\documentclass[conference]{IEEEtran}
\IEEEoverridecommandlockouts
\pdfoutput=1
\usepackage{cite}
\usepackage{amsmath,amssymb,amsfonts}
\usepackage{algorithmic}
\usepackage{graphicx}
\usepackage{textcomp}
\usepackage{array}
\usepackage{xcolor}
\usepackage{comment}
\usepackage{caption}
\usepackage{subcaption}
\def\BibTeX{{\rm B\kern-.05em{\sc i\kern-.025em b}\kern-.08em
    T\kern-.1667em\lower.7ex\hbox{E}\kern-.125emX}}
\begin{document}

\title{Data Diversity based Metamorphic Relation Prioritization}

\author{\IEEEauthorblockN{ Madhusudan Srinivasan}
\IEEEauthorblockA{\textit{Montana State University}\\
Bozeman, United States \\
madhusuda.srinivasan@montana.edu}
\and
\IEEEauthorblockN{{Upulee Kanewala}
\IEEEauthorblockA{\textit{University of North Texas}\\
Florida, United States \\
upulee.kanewala@unf.edu}
}}
\maketitle

\begin{abstract}
 An oracle is a mechanism to decide whether the outputs of the program for the executed test cases are correct. For machine learning programs, such oracle is not available or too difficult to apply. Metamorphic testing is a testing approach that uses metamorphic relations, which are necessary properties of the software under test to help verify the correctness of a program. Prioritization of metamorphic relations helps to reduce the cost of metamorphic testing~\cite{anonymous2021}.  However, prioritizing metamorphic relations based on code coverage is often not effective for prioritizing MRs for machine learning programs, since the decision logic of a machine learning model is learned from training data, and 100\% code coverage can be easily achieved with a single test input. To this end, in this work, we propose a cost-effective approach based on diversity in the source and follow-up data set to prioritize metamorphic relations for machine learning programs. We show that the proposed data diversity-based prioritization approach increase the fault detection effectiveness by up to 40\% when compared to the code coverage-based approach and reduce the time taken to detect a fault by 29\% when compared to random execution of MRs. Overall, our approach leads to saving time and cost during testing.
\end{abstract}

\begin{IEEEkeywords}
Metamorphic Testing, Metamorphic Relations, Machine Learning
\end{IEEEkeywords}
\section{Introduction and Motivation}

Machine learning (ML) is increasingly deployed in large-scale software and safety-critical systems due to recent advancements in deep learning and reinforcement learning. The software applications powered by ML are applied and utilized in our daily lives; from finance, and energy, to health and transportation. AI-related accidents are already making headlines, from inaccurate
facial recognition systems causing false arrests to unexpected racial and gender discrimination by machine learning software~\cite{aincident}~\cite{googlecar}~\cite{aincident1}. Also, a recent incident with an Uber autonomous car resulted in the death of a pedestrian~\cite{UberSelfDriving}. Thus, ensuring the reliability and conducting rigorous testing of machine learning systems is very vital to prevent any future disasters.


One of the critical components of software testing is the test oracle.
The mechanism to determine the correctness of test case output is known as test oracle~\cite{barr2015oracle}. A program can be considered a non-testable program if there does not exist a test oracle or too difficult to determine whether the test output is correct~\cite{weyuker1982testing}. Hence, those programs suffer from a problem known as the test oracle problem. ML-based systems exhibit non-deterministic behavior and reason in probabilistic manner. Also, the output is learned and predicted by a machine learning model as result do not have an expected output to compare against the actual output. As a result, correctness of the output produced by machine learning-based applications cannot be easily determined and suffer from test oracle problem~\cite{zhang2020machine}.  

Metamorphic testing (MT) is a property-based software testing technique that is used to alleviate the test oracle problem. A central component of MT is a set of Metamorphic Relations (MRs), which are necessary properties of the target function or algorithm in relation to multiple inputs and their expected
outputs~\cite{chen2018metamorphic}~\cite{segura2018metamorphic}. In recent years especially, MT has attracted a lot of attention
and detected a large number of real-life faults in various domains. For example, Murphy et al~\cite{murphy2008properties} identified six metamorphic relations and real bugs were identified on three machine learning tools. In a related
project, Xie et al~\cite{xie2011testing}, \cite{xie2009application} used metamorphic
testing for the detection of faults in K-nearest neighbours and Naive Bayes classifiers. The results
revealed that both algorithms violated some of the metamorphic relations. Also, some real faults were
detected in the open–source machine learning tool Weka~\cite{gewehr2007bioweka}. Zhang et al~\cite{zhang2018deeproad} applied MT to test DNN based autonomous driving systems. MT was able to find
inconsistent behaviors under different road scenes for real-world autonomous driving models.

Previous work~\cite{Anonymous2018} has shown that several MRs detect the same type of fault and differ only in fault detection effectiveness. Also, each MR in a machine learning application can have multiple source and follow-up test cases, as a result, the execution time of the MRs increases drastically.  The source and follow-up test cases in a MR represented as training data set could take several days to train and build a machine learning model for a large training data set. Hence, prioritization of MRs is important for conducting effective metamorphic testing on these programs. 

Prior work~\cite{anonymous2021} introduced
approaches to prioritize MRs based on code coverage and fault-based approach. The fault-based prioritization approach involves using mutant killed information of each MR to prioritize MR. The code coverage approach uses statement or branch coverage information for prioritizing MRs. Further, the code coverage-based approach does not work effectively for machine learning-based applications since the data used for training determines the program logic. Machine learning program is usually just a sequence of library function invocations and as a result 100\% code coverage can be easily achieved when at least one test input is processed. To this end, in this work, we propose methods for \emph{MR prioritization} for machine learning programs. The MR prioritization method proposed in this work qualitatively identifies the diversity between the source and follow-up training data set in a MR. Moreover, our results indicate that MR prioritization reduces the number of MRs needed to execute on SUT and reduces the time taken to detect a fault up to 29\%. 

We make the following contributions in this work:
\begin{enumerate}
    \item Propose a novel data diversity based automated method where we proposed four metrics to prioritize \emph{metamorphic relations} based on diversity between the source and follow-up data set in a MR.
    \item  Evaluate the effectiveness of our proposed MR prioritization methods on four machine learning algorithm implementations and perform a comparison against two baseline approaches (1) Random baseline: represents the current practice of executing the MRs in a random order, 
    (2) Code coverage-based ordering: represents the MR ordering based on the statement or branch coverage information collected when testing the previous version of the SUT as show in the prior work~\cite{anonymous2021}. 
            \end{enumerate}
            
\section{Proposed Approach}
\label{sec:mlmetrics}
\begin{figure*}[h]
\centering
\includegraphics[width=0.85\textwidth]{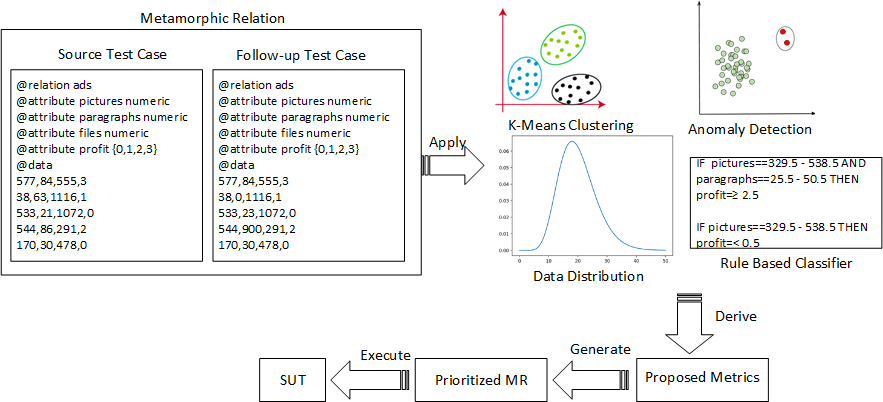}
\caption{Steps for MR prioritization of ML Programs}
\label{label:ML_Approach}
\end{figure*}
In this section, we discuss our proposed method for prioritization of MRs. Figure~\ref{label:ML_Approach} shows the steps for prioritizing MRs. 
In this approach, we  prioritize the MRs as follows: 
\begin{enumerate}
\item Let the set of source test cases in a MR used for testing the SUT be the \emph{prioritizing source test cases} $(T_{sp})$.
    \item Let the set of follow-up test cases in a MR used for testing the SUT be the \emph{prioritizing follow-up test cases} $(T_{fp})$. 
 \item For $(T_{sp})$ and $(T_{fp})$, apply the proposed metric: rule based, anomaly detection, clustering based and data distribution. The metrics are discussed in detail in section \ref{sec:datadiversity_approach}.
  \item For $(T_{sp})$ and $(T_{fp})$, apply the proposed metric: rule based, anomaly detection, clustering based and data distribution and get the MR diversity value of each MR based on the formulas provided in section \ref{sec:datadiversity_approach}.
  \item MR diversity value calculated using the metrics for each MR is normalized using the following steps:
  \begin{enumerate}
  \item Let the set of MR diversity value of MRs be \emph{$DMR_{v}$}.
  \item Identify minimum and maximum value in $DMR_{v}$. Let minimum value be $Min_{v}$ and maximum value be $Max_{v}$.
  \item For each MR diversity value $v\in DMR_{v}$, generate the normalized MR diversity value using the formula below:
  \begin{equation}
 N_{MR}=\frac{v-Min_{v}}{Max_{v}-Min_{v}}
\label{label:normalizeMRvalue}
\end{equation}
Where $N_{MR}$ represent the normalized MR diversity value.
\end{enumerate}
       \item Prioritize the normalized MR diversity values and select the top \emph{n} MRs from the prioritized ordering to execute based on the resources available for testing.
       \end{enumerate}
Our proposed metrics are explained in detail below.

\subsection{Dataset Diversity Approach}
\label{sec:datadiversity_approach}


In this work, we propose four metrics to prioritize metamorphic relations based on diversity between source and follow-up data set in a MR. Our intuition behind the metrics is that greater the diversity between the source and follow-up data set, then greater the quality of MR and performance of the ML model.

The proposed metrics are explained in detail below.

\subsubsection{Rule Based Classifier}
Rule-based classifiers are just another type of classifier which makes the class decision by using various if-then rules. We apply a rule based classifier (CN2) to determine diversity between the source and follow-up data set in a metamorphic relation. CN2 is an instance of sequential covering algorithm family based on Rule-Based classification used for learning set of rules.The basic idea here is to learn one rule, remove the data that it covers, then repeat the same process. We hypothesize that if greater the diversity in the classification rules between the source and followup data set, then greater the performance of the model and MR. We apply the following steps.
\begin{enumerate}
 \item Let the set of source test cases used for testing SUT  be the \emph{prioritizing source test cases} $(T_{sp})$.
     \item Let the set of follow-up test cases of the SUT be the \emph{prioritizing follow-up test cases} $(T_{fp})$.
\item Apply Sequential Covering Algorithm (CN2) on $T_{sp}$ and $T_{fp}$ and obtain the rules.The CN2 induction algorithm is a learning algorithm for induction of simple and comprehensible rules.
\item Remove same set of rules covered in both $T_{sp}$ and $T_{fp}$.
\item Calculate the total rules identified in $T_{sp}$ and $T_{fp}$. Let $Total_{sr}$ and $Total_{fr}$ represent the total rules in $T_{sp}$ and $T_{fp}$ respectively.

\item We calculate the diversity of MR using the formula below:
\begin{equation}
\begin{split}
\text{$R_{MR}$} =\text{$\mid(Total_{sr})-(Total_{fr}) \mid$}
\end{split}
\label{label:rulebased}
\end{equation}
Where $R_{MR}$ represent the diversity of the MR.
\end{enumerate}

\subsubsection{Anomaly Detection}
Anomaly detection (or outlier detection) is the identification of rare items, events or observations which raise suspicions by differing significantly from the majority of the data. We apply anomaly detection to identify outliers in the source and followup data set in a metamorphic relation. We hypothesize that greater the diversity in outliers detected between the source and followup, greater the possibility of detecting fault in the machine learning model. We apply the following steps.

\begin{enumerate}
 \item Let the set of source test cases used for testing SUT  be the \emph{prioritizing source test cases} $(T_{sp})$.
     \item Let the set of follow-up test cases of the SUT be the \emph{prioritizing follow-up test cases} $(T_{fp})$.
\item Apply K Nearest Neighbour (KNN) algorithm to $T_{sp}$ and $T_{fp}$ and identify outlier instances.
\item Remove identical outliers detected between $T_{sp}$ and $T_{fp}$.
\item Calculate the total outliers in $T_{sp}$ and $T_{fp}$ respectively. Let $Total_{so}$ and $Total_{fo}$ be the sum of outlier detected in $T_{sp}$ and $T_{fp}$ respectively.
\item We calculate the diversity of MR using the formula below:
\begin{equation}
\begin{split}
\text{$O_{MR}$} =\text{$\mid(Total_{so})-(Total_{fo}) \mid$}
\end{split}
\label{label:anomalydetection}
\end{equation}
Where $O_{MR}$ represent the diversity of the MR.
\end{enumerate}


\subsubsection{Clustering-based}
In this metric, we partition the input space ie. all possible inputs into different region and determine how the data is distributed in the space. To achieve this, we use clustering methods which simply try to group similar patterns into clusters whose members are more similar to each other based on some distance measure than to members of other clusters. We apply the following steps.

\begin{enumerate}
 \item Let the set of source test cases used for testing SUT  be the \emph{prioritizing source test cases} $(T_{sp})$.
     \item Let the set of follow-up test cases of the SUT be the \emph{prioritizing follow-up test cases} $(T_{fp})$.
\item Apply K mean algorithm to $T_{sp}$ and $T_{fp}$ and identify clusters. Let clusters of $T_{sp}$ and $T_{fp}$ be $S_{cl}$ and $F_{cl}$.
\item Find the distance between the clusters by calculating the euclidean distances between the clusters in $S_{cl}$. Let the total distance between the clusters be $Ts_{cds}$.
\item Determine the size of each cluster in  $S_{cl}$. Let the total size of the clusters in $S_{cl}$ be $Ts_{sc}$.
\item Find the distance within the cluster by calculating the average of the distances from observations to the centroid of each cluster in $S_{cl}$. Let average distance within the cluster in $S_{cl}$ be $As_{cd}$.
\item Apply the steps 4 to 6 to $F_{cl}$ respectively. Let $Tf_{cds}$ be the total distance between the clusters, $Tf_{sc}$ be the total size of the clusters and $Af_{cd}$ be the average distance within the cluster. 

\begin{equation}
\begin{split}
\text{$R_{MR}$} =\text{$\mid(Total_{sr})-(Total_{fr}) \mid$}
\end{split}
\label{label:rulebased}
\end{equation}

\item We calculate the diversity of MR using the formula below:
\begin{equation}
\begin{split}
\begin{split}
\text{$C_{MR}$} 
=\text{$\mid(Ts_{cds}+Ts_{sc}+As_{cd})-(Tf_{cds}+Tf_{sc}+Af_{cd})\mid$}
\end{split}
\end{split}
\label{label:anomalydetection}
\end{equation}
Where $C_{MR}$ is the diversity of MR.
\end{enumerate}

\subsubsection{Data Distribution}
Data used in training machine learning model often form very similar patterns. For example, a lot of data were grouped around the middle values, with fewer observations at the outside edges of the distribution. These patterns are known as distributions, because they describe how the data are distributed across the range of possible values. In this metric, we used shape and spread of data distribution of dataset to find diversity of metamorphic relations. The spread of distribution involves calculating range, variance and standard deviation of the source and follow-up data set. Similarly, shape of distribution involves calculating skewness and kurtosis of the source and follow-up data set of the MR. We hypothesize that larger the diversity in the data distribution between the source and follow-up data set of a  MR, then greater the quality of MR. We apply the following steps.
\begin{enumerate}
 \item Let the set of source test cases used for testing SUT  be the \emph{prioritizing source test cases} $(T_{sp})$. 
     \item Let the set of follow-up test cases of the SUT be the \emph{prioritizing follow-up test cases} $(T_{fp})$. 
     \item \label{steps3:datadis} Calculate the skewness and kurtosis for $T_{sp}$. Let $ST_{sk}$ be the sum of skewness and kurtosis for $T_{sp}$.
     \item \label{steps4:datadis} Calculate range, variance and standard deviation for $T_{sp}$. Let $ST_{rvs}$ be the sum of variance, range and standard deviation for $T_{sp}$.
     \item Apply step \ref{steps3:datadis} and {\ref{steps4:datadis}} to $T_{fp}$. Let $FT_{sk}$ and $FT_{rvs}$ be the shape and spread of distribution for $T_{fp}$.
     
     \item We calculate the diversity of MR using the formula below:
\begin{equation}
\begin{split}
\text{$DD_{MR}$}
\begin{split}
=\text{$\mid(ST_{sk}+ST_{rvs})-(FT_{sk}+FT_{rvs})\mid$}
\end{split}
\end{split}
\label{label:anomalydetection}
\end{equation}
Where $DD_{MR}$ is the diversity of MR.
\end{enumerate}

\section{Evaluation}
\label{sec:eval}
In this work, we plan to evaluate the utility of the developed prioritization approaches on the following aspects: (1) Fault detection effectiveness of MR sets, (2) Effective number of MRs required for testing, and (3) Time taken to detect a fault. We evaluate the effectiveness of the proposed approaches with (1) a \emph{random baseline}: this represents the current practice of executing source and follow-up data set of the MRs in random order.
(2) \emph{code-coverage based ordering}: this represents the MR order based on the statement and branch-coverage information of the MRs. The approach selects MR which has covered highest number of statements or branches and place it in the prioritized MR ordering. The process continues until all the possible statement or branches are covered~\cite{anonymous2021}.

We conducted experiments to find answers for the following research questions:
\begin{enumerate}
    \item Research Question 1 (RQ1): \emph{Are the proposed MR prioritization approach based on data diversity more effective than the random baseline?}
    \item Research Question 1 (RQ2): \emph{Are the proposed MR prioritization approach more effective than the code coverage-based prioritization?}
    \end{enumerate}


\subsection{Evaluation Procedure}
\label{evalProc_method2}
In order to answer the above research questions, we carry out the following validation procedure similar to previous work~\cite{anonymous2021}:

\begin{enumerate}

\item We generate a set of mutants for the SUT. These mutants will be referred to as the \emph{validation set of faults} $(F_v)$. Then from the generated mutants, we remove mutants that give exceptions or mutants that do not finish execution within a reasonable amount of time from further consideration.

\item We use the method described in Section~\ref{sec:mlmetrics} to obtain data diversity-based MR ordering. Then we applied the obtained MR ordering to $F_v$ and log the mutant killing information.

\item \textit{Creating the random baseline}: We generate 100 random MR orderings and apply each of those orderings to $F_v$. The mutant killing information for each of those random orderings is logged. The average mutant killing rates of these 100 random orderings were computed to get the fault detection effectiveness of the random baseline.

\item \textit{Creating the Code Coverage-based ordering}: Apply greedy approach on the statements or branches covered to prioritize the MRs as follows~\cite{anonymous2021}:
\begin{enumerate}
\item Select the MR that covers the highest number of statement or branches in the program and place it in the prioritized MR ordering.
\item Remove the statements and branches covered by the MR.
\item Repeat the previous two steps until all the statements and branches are covered.
\end{enumerate}

\end{enumerate}

\subsection{Evaluation Measures}
\label{evalMeasures}
We used the following measures to evaluate the effectiveness of the MR orderings generated by our proposed metrics:
\begin{enumerate}
\item To measure the \emph{fault detection effectiveness} of a given set of MRs, we use the percentage of mutants killed by those MRs.
        \item To calculate the \emph{effective MR set size} we used the following approach: typically, the fault detection effectiveness of  MT  increases  as the number of MRs used for testing increases. However, after a certain number of MRs, the rate of increase in fault detection slows due to factors such as redundancy of MRs. Therefore when there is no significant increase in the fault detection between two MR sets of consecutive sizes of size $m$ and $m+1$, where the MR set of size $m+1$ is created by adding one MR to the MR set set of size $m$, the effective MR set size can be determined. That is, if the difference in fault detection effectiveness of MR set of size $m$ and MR set of size $m+1$ is less than some threshold value, $m$ would be the effective MR set size that should be used for testing. Determining this threshold value should be done considering the critical nature of the SUT. In this work, we used two threshold values of 5\% and 2.5\% as used in previous related work for determining the oracle data set size~\cite{gay2015automated}.
    \item We used the following approach to find the \emph{average time taken to detect a fault}: for each killable mutant $m$ in $F_v$, we computed the time taken to kill the mutant ($t_m$) by computing the time taken to execute the source and follow-up test cases of the MRs in a given prioritized order until $m$ is killed (here it is assumed that the source and follow-up test cases for each MR are executed sequentially). Then the average time taken to detect a fault is computed using the following formula: $$\frac{\sum t_m}{\#killable\, mutants\, in\, F_v}$$
    \item We calculate \emph{Average percentage of fault detected} (APFD) developed by Elbaum et al.~\cite{elbaum2002test}~\cite{malishevsky2006cost}~\cite{elbaum2004selecting} that measures the average rate of fault detection per percentage of test suite execution. The APFD is calculated by taking the weighted average of the number of faults detected during the execution of MRs. APFD can be calculated using the formula below: 
    \begin{equation}
\begin{split}
\text{APFD} =
 \frac{1-MRF_1+MRF_2+...+MRF_m}{nm}+\frac{1}{2n}
\end{split}
\label{label:apfd_formula}
\end{equation}
Where $MRF_i$ represents the fault $F_i$ detected by the metamorphic relations under evaluation, m represents the number of faults present in the SUT and n represents the total number of MRs used.

\end{enumerate}

\subsection{Subject Programs and MRs}
    
   We applied the above-mentioned validation procedure on the following four machine learning programs for evaluating our proposed MR prioritization methods. 
    \begin{itemize}
        \item IBk\footnote{http://weka.sourceforge.net/doc.dev/weka/classifiers/lazy/IBk.html}: K-nearest neighbours (KNN) classifier implementation  in the Weka machine learning library~\cite{Aha1991}. The input to IBk is a training data set,  and a test data set to represent in .arff format. The output is the classification predictions made on the instances in the test data set. 
        \item Linear Regression\footnote{https://weka.sourceforge.io/doc.dev/weka/classifiers/functions/LinearRegression}: linear regression is a linear approach that models the relationship between the input variables and the single output variable. The subject program is the linear regression implementation in weka library. Weka latest stable version is 3.8.5, which consists of the multiple linear regression, named \textit{LinearRegression()} class. The input to linear regression program is a training data set,  and a test data set to represent in .arff format. The output is the prediction made on the instances in the test data set.
        \item Naive Bayes\footnote{https://weka.sourceforge.io/doc.dev/weka/classifiers/bayes/NaiveBayes.html}: A Naive Bayes classifier is a probabilistic machine learning model that’s used for classification task. The subject program is the Naive Bayes algorithm implementation, named \textit{NaiveBayes()} class in Weka library. The input to Naive Bayes program is a training data set,  and a test data set to represent in .arff format. The output is a classification made on the instances in the test data set.
        \item Convolutional Neural Network\footnote{https://javadoc.io/doc/org.deeplearning4j/deeplearning4j-nn/latest/index.html}: Convolutional Neural Network (CNN) is a class of Artificial Neural Network and most commonly applied to analyze visual imagery. The subject program is the CNN algorithm implementation in Deeplearning4j library. The input to the CNN is MNIST handwritten digit data set and the output is to classify the image of a handwritten digit into one of 10 classes representing integer values from 0 to 9.
    \end{itemize}
    
\subsection{Metamorphic Relations}
For conducting MT on IBk and Naive Bayes, we used 11 MRs developed by Xie et al.~\cite{xie2011testing}. These MRs are developed based on the user expectations of supervised classifiers. These MRs modify the training and testing data so that the predictions do not change between the source and follow-up test cases. Similarly, for testing Linear Regression system, we used 11 MRs developed by Luu et al.~\cite{luu2021testing}. The properties of the linear regression is related to addition of data points, the rescaling of inputs, the shifting of variables, the reordering of data, and the rotation of independent variables. These MRs were grouped into 6 categories and derived from the properties of the targeted algorithm. For testing convolutional neural network, we used 11 MRs developed by Ding et al.  ~\cite{ding2017validating}. The MRs were developed on three levels: system level, data set level and data item level. The MRs in the data set level are created
on the relation of the classification accuracy of reorganized training data sets. The MRs in the data item level are created on the relation of the classification accuracy of reproduced
individual images.


\subsection{Source and Follow-up Test Cases}
\label{testcases}

As we described before, MT involves the process of generating source and follow-up test cases based on the MRs used for testing.  

IBk uses a training data set to train the k-nearest neighbor classifier, and a test data set is used to evaluate the performance of the trained classifier. After executing source test case, the output will be the class labels predicted for the test data set, which is a value from the set \{0,1,2,3,4\}. The attribute values are randomly selected within the range [0, 100]. The values of the class label are randomly selected from the set \{0,1,2,3,4,5\}. The size of the training testing data sets ranges within [0, 500]. To generate the follow-up test cases using these source test cases, we applied the input transformations described in the MRs. 

Linear Regression uses a training data set to train the model, and a test data set is used to evaluate the performance of the trained model. We obtained the training data from machine learning repository \footnote{https://archive.ics.uci.edu/ml/datasets/YearPredictionMSD} as source data set. The training data contains 515345 instances and 90 attributes. To generate the follow-up data set using these source data set, we applied the input transformations described in the MRs. Similarly for Naive Bayes, we obtained the training data from machine learning repository \footnote{https://archive.ics.uci.edu/ml/datasets/Adult} as source data set. The training data contains 48842 instances and 14 attributes. To generate the follow-up data set using these source data set, we applied the input transformations described in the MRs. For testing convolutional neural network, we obtained the training data~\footnote{http://yann.lecun.com/exdb/mnist/} as source data set. The training dataset contains 60000 handwritten digits and 10000 images for testing. To generate the follow-up test cases using these source test cases, we applied the input transformations described in the MRs.

\subsection{Mutant Generation}
\label{sec:mutants}
For each subject program, we aimed at developing generated mutant set to be used as $F_v$. For this, we used three automated mutation tools: $\mu$Java\footnote{https://github.com/jeffoffutt/muJava}, PIT\footnote{http://pitest.org/} and Major\footnote{http://mutation-testing.org/}. For generating mutants using $\mu$Java, we used all the method level mutation operators~\cite{ma2006mujava}.  With PIT mutation tool~\cite{coles2016pit} and the Major mutation tool~\cite{just2014major} we used all the available mutation operators provided by the tools. All the mutants the were generated using these tools were \emph{first order mutants}, where a mutation operator is used to make a single modification to the source code to generate the mutant. Table~\ref{Tab:Mutants_vk_method2} shows the number of mutants used for evaluation.

 .

\begin{table}[h]
\centering
\caption{Mutants generated for version $v_{k}$ of the SUTs}
\label{Tab:Mutants_vk_method2}
\begin{tabular}{|c|c|c|c|c|}
\hline
\textbf{Version} & \textbf{Subject} & \textbf{\#Major} & \textbf{\#PIT} & \textbf{\# $\mathbf{\mu}$Java} \\ \hline
3.8.6            & IBk              & 1021               & N/A             & 621             \\ \hline
3.8.6            & Linear Regression         & 377              &   73         & N/A             \\ \hline
3.8.6            & Naive Bayes         & 452              & 452            & N/A             \\ \hline
0.9.1           & CNN         & 92              & N/A            & N/A             \\ \hline
\end{tabular}
\end{table}



\begin{table}[h]
\centering
    \caption{Validation Setup}
    \label{evalSetup_method2}
\begin{tabular}{|c|c|c|}
\hline
\textbf{Subject}  & \textbf{$\mathbf{{T_{sp}}}$}   & \textbf{$\mathbf{{F_v}}$} \\ \hline
IBk               & Test1          & Major       \\ \hline
Linear Regression & YearPrediction & Major       \\ \hline
Naive Bayes       & Adult          & PIT         \\ \hline
CNN               & MNIST          & PIT         \\ \hline
\end{tabular}
\end{table}


\section{Result and Analysis}
\label{section:Results&Discussion}
\begin{figure*}[!ht]
\centering

~\hfill
   \begin{subfigure}[b]{0.49\textwidth}
      \includegraphics[width=\textwidth]{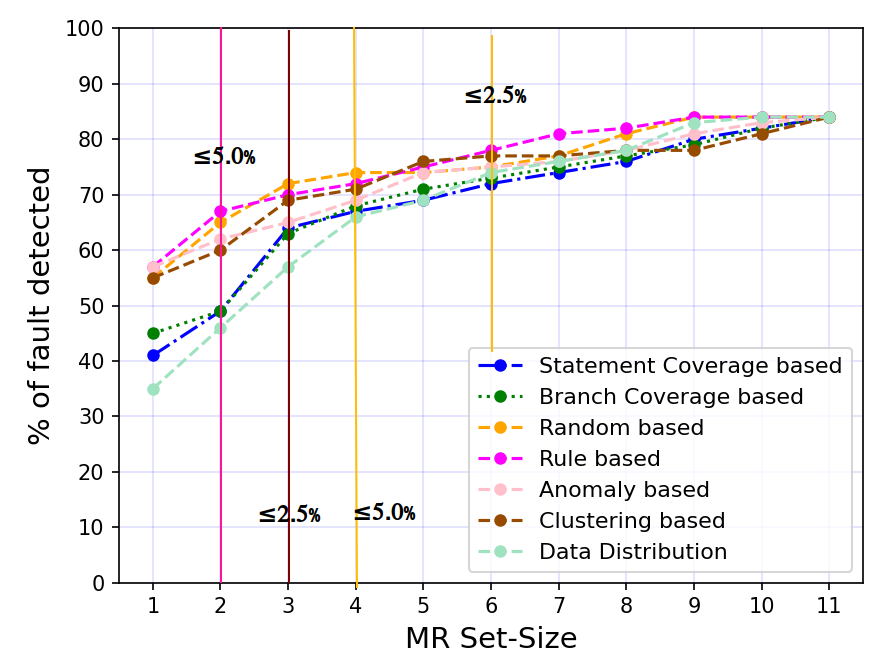}
    \caption{IBk}
    \label{fig:IBk_method_killrateMRs}
  \end{subfigure}
  \\
       \begin{subfigure}[b]{0.49\textwidth}
      \includegraphics[width=\textwidth]{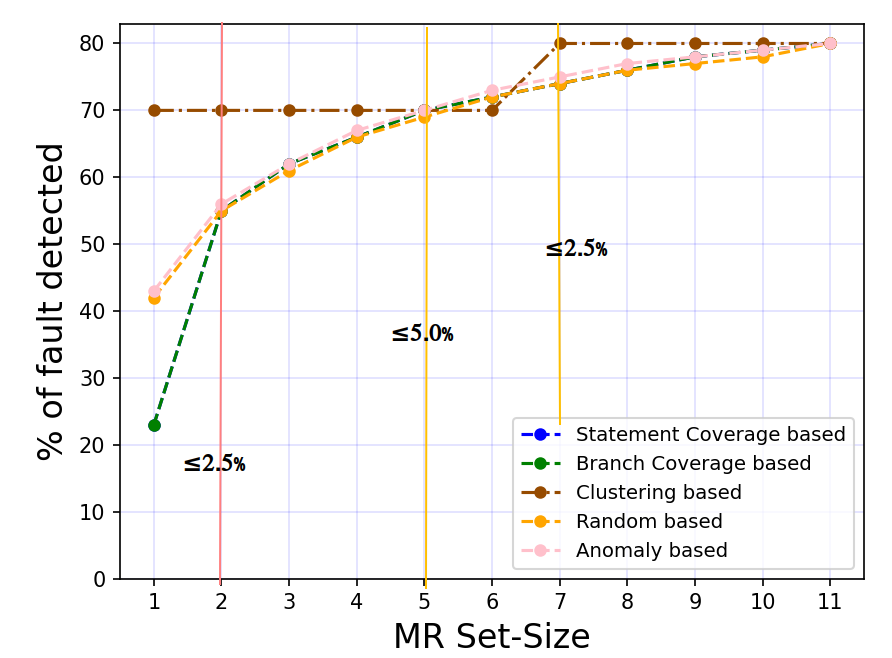}
    \caption{CNN}
    \label{fig:cnn_method2_killrateMRs}
  \end{subfigure}
  \hfill
  \begin{subfigure}[b]{0.49\textwidth}
      \includegraphics[width=\textwidth]{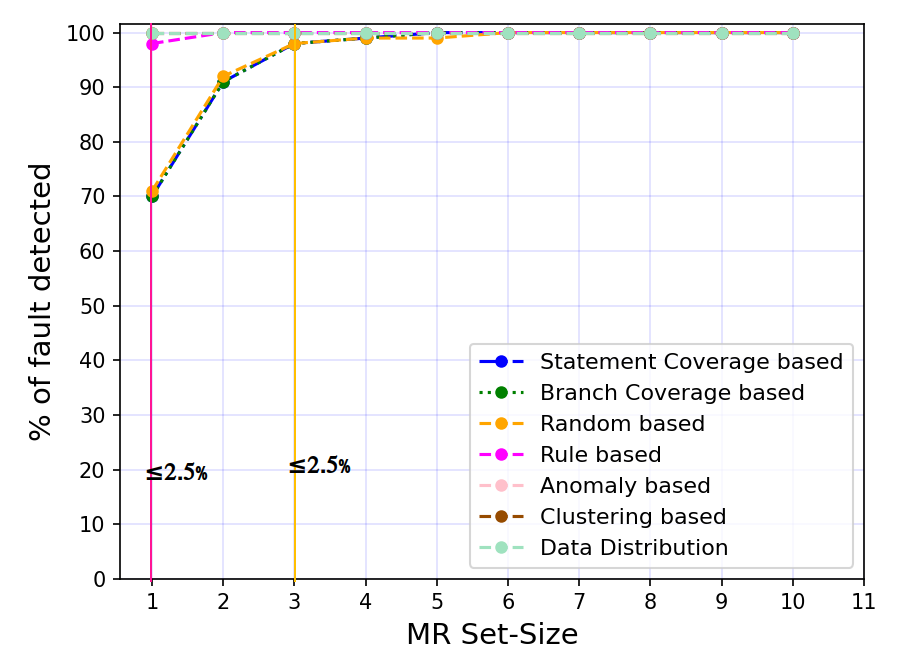}
    \caption{Naive Bayes}
    \label{fig:naivebayes_method2_killrateMRs}
  \end{subfigure}
  \hfill
  \begin{subfigure}[b]{0.49\textwidth}
      \includegraphics[width=\textwidth]{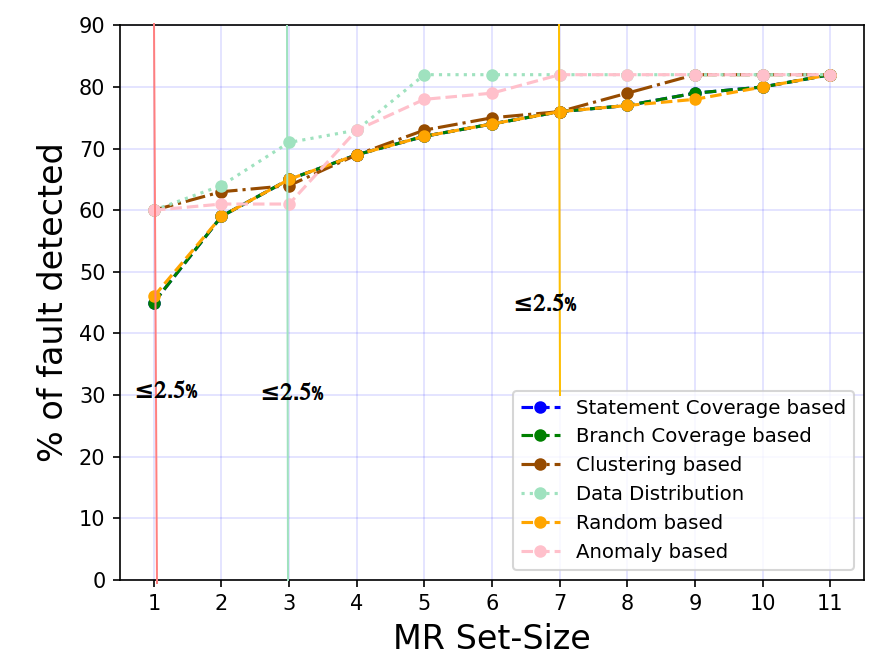}
    \caption{Linear Regression}
    \label{fig:regression_method2_killrateMRs}
  \end{subfigure}
  
     \caption{Fault detection effectiveness for IBk, Linear Regression, Naive Bayes and Convolutional Neural Network }
    \label{Fig:allsubject_methods_killrateMRs}
\end{figure*}

In this section, we discuss our experimental results and provide answers to the two research questions that we listed in Section~\ref{sec:eval}.
For each subject program, we carried out the validation procedure described in Section~\ref{evalProc_method2} using the setup described in Table~\ref{evalSetup_method2}. In this setup, we used the generated mutant set for evaluating the prioritized MR ordering as $F_v,$ and the source test cases as $T_{sp}$. For example, in Table~\ref{evalSetup_method2} for IBk, Test1 refers to the randomly generated data set that we described in Section~\ref{testcases}. Similarly, for CNN, MNIST refers to the 60000 images used for training and testing. Adult and YearPrediction refers to the training data set used for Naive Bayes and Linear Regression program respectively. Figure~\ref{Fig:allsubject_methods_killrateMRs} shows the average fault detection effectiveness for evaluation runs described above vs. the MR set size used for testing, for each subject program. We also plot the percentage of faults detected by the random baseline, statement and branch coverage based MR ordering for comparison. Results for each research question is provided below.

\subsection{RQ1: Comparison of the MR Prioritization approaches against the Random Baseline}

\subsubsection{Fault detection effectiveness}

We formulated the following statistical hypothesis to answer RQ1 in the context of fault detection effectiveness:
\begin{itemize}
\item[$H_{1}$:]For a given MR set of size \textit{m}, the fault detection effectiveness of the MR set produced by data diversity-based approach is higher than that of the random baseline.

\end{itemize}
The null hypothesis $H_{0x}$ for each of the above defined hypothesis $H_{X}$ is that the data-diversity based approaches perform equal to or worse than the random baseline.

In Table~\ref{tab:RL_method3_random_IBK} to~\ref{tab:RL_method3_random_CNN}, we list the relative improvement in average fault detection effectiveness for the data diversity-based method over the currently used random approach of prioritizing MRs for Linear Regression, Naive Bayes, IBk and CNN respectively. To evaluate the above hypotheses, we used the paired permutation test since it does not make any assumptions about the underlying distribution of the data~\cite{good2013permutation}. We apply the paired permutation test to each MR set size for each of the subject programs with $\alpha$= 0.05, and the relative improvements that are significant are marked with a *. 

As shown in Table~\ref{tab:RL_method3_random_IBK}, for IBk, rule based metric shows improvements in average fault detection effectiveness over the random approach for all the MR set sizes except for the last MR set size. Particularly, among the different MR set sizes, the increase in fault detection percentage varies from 1.23\% to 29.5\%.  Therefore, we reject the null hypotheses $H_{01}$ for IBk for rule based, Anomaly, clustering based and data distribution metric. Similarly for Linear Regression, Table~\ref{tab:RL_method3_random_Regression} indicates that the anomaly based, clustering based and data distribution based metrics show consistent significant improvements in fault detection over random. Therefore we can reject $H_{01}$ in general for Regression. The fault detection varies from 2.5\% to 30\% for the three metrics. Table~\ref{tab:RL_method3_random_Naive} shows the relative improvement in the average fault detection percentage for Naive Bayes. Our proposed metrics shows improvements for MR set sizes $m=1$ to $m=5$. Also overall, the relative improvement across MR prioritization methods varies between 0\% - 40\%. Therefore we reject $H_{01}$ in general for Naive Bayes.
Table~\ref{tab:RL_method3_random_CNN} shows the relative improvement in the average fault detection percentage for CNN. Clustering-based metric shows improvements for all MR set sizes except MR set size $m=5$. Also overall, the relative improvement across MR prioritization methods varies between 0\% -66\%. Therefore we can reject $H_{01}$ in general for CNN.

Therefore our data diversity based approach outperforms the random approach for all our subject programs.

\begin{table}[h]
\centering
 \caption{Relative improvement in fault detection effectiveness of the data diversity-based approach compared to random baseline for IBk}
     \label{tab:RL_method3_random_IBK}
        \begin{tabular}{|>{\centering\arraybackslash}p{1.2cm}|>{\centering\arraybackslash}p{1.2cm}|>{\centering\arraybackslash}p{1.2cm}|>{\centering\arraybackslash}p{1.2cm}|>{\centering\arraybackslash}p{1.2cm}|}
\hline
\textbf{MR Set-Size} & \textbf{Rule-based} & \textbf{Anomaly-based} & \textbf{Clustering-based} & \textbf{Data Distribution} \\ \hline
1                    & 29.54\%*               & 29.54\%*                  & 25\%*                        & 20.45\%*                     \\ \hline
2                    & 17.54\%*               & 8.77\%*                   & 5.26\%*                      & 19.29\%*                     \\ \hline
3                    & 11.11\%*               & 3.17\%*                   & 9.52\%*                      & 9.52\%*                      \\ \hline
4                    & 5.88\%*                & 1.47\%*                   & 4.41\%*                      & 2.94\%*                      \\ \hline
5                    & 5.63\%*                & 4.22\%*                   & 7.04\%*                      & 2.81\%*                      \\ \hline
6                    & 5.40\%*                & 1.35\%*                   & 4.05\%*                      & 0\%                          \\ \hline
7                    & 6.57\%*                & 0\%                      & 1.31\%*                      & 0\%                          \\ \hline
8                    & 5.12\%*                & 0\%                      & 0\%                         & 0\%                          \\ \hline
9                    & 5\%*                   & 1.25\%*                   & -2.5\%                      & 3.75\%*                       \\ \hline
10                   & 2.43\%*     & 1.21*       & -1.21\%         & 2.43\%*            \\ \hline
11                   & 0\%                   & 0\%                      & 0\%                         & 0\%                          \\ \hline
\end{tabular}
\end{table}

\begin{table}[h]
\centering
\caption{Relative improvement in fault detection effectiveness of the data diversity based approaches compared to random baseline for Linear Regression}
\label{tab:RL_method3_random_Regression}
\begin{tabular}{|c|c|c|c|}
\hline
\textbf{MR Set-Size} & \textbf{Anomaly-based} & \textbf{Clustering-based} & \textbf{Data Distribution} \\ \hline
1                    & 30.43\%*                  & 30.43\%*                      & 30.43\%*                      \\ \hline
2                    & 3.38\%*                   & 6.77\%*                     & 8.47\%*                       \\ \hline
3                    & -6.15\%                   & 1.53\%*                        & 9.23\%*                          \\ \hline
4                    & 5.79\%*                  & 1.44\%*                     & 5.79\%*                       \\ \hline
5                    & 8.33\%*                  & 1.38\%*                     & 13.88\%*                       \\ \hline
6                    & 6.75\%*                      & 1.35\%*                    & 10.81\%*                       \\ \hline
7                    & 7.89\%*                  & 0\%*                      & 7.89\%*                       \\ \hline
8                    & 6.49\%*                      & 2.59\%*                         & 6.49\%*                          \\ \hline
9                    & 5.12\%*                   & 5.12\%*                      & 5.12\%*                       \\ \hline
10                   & 2.5\%*                   & 2.5\%*                      & 2.5\%*                       \\ \hline
11                   & 0\%                      & 0\%                         & 0\%                          \\ \hline
\end{tabular}
\end{table}

\begin{table}[h]
\centering
 \caption{Relative improvement in fault detection effectiveness of the data diversity-based approach compared to random baseline for Naive Bayes}
     \label{tab:RL_method3_random_Naive}
     \begin{tabular}{|>{\centering\arraybackslash}p{1.4cm}|>{\centering\arraybackslash}p{1.4cm}|>{\centering\arraybackslash}p{1.4cm}|>{\centering\arraybackslash}p{1.4cm}|>{\centering\arraybackslash}p{1.4cm}|}
      
\hline
\textbf{MR Set-Size} & \textbf{Rule-based} & \textbf{Anomaly-based} & \textbf{Clustering-based} & \textbf{Data Distribution} \\ \hline
1                    & 38.02\%*               & 40.84\%*                  & 40.84\%*                     & 40.84*                      \\ \hline
2                    & 8.69\%*                & 8.69\%*                   & 8.69\%*                      & 8.69*                       \\ \hline
3                    & 2.04\%*                & 2.04\%*                  & 2.04\%*                      & 2.04\%*                       \\ \hline
4                    & 1.01\%*                & 1.01\%*                   & 1.01\%*                      & 1.01*                       \\ \hline
5                    & 1.01\%*                & 1.01\%*                   & 1.01\%*                      & 1.01*                       \\ \hline
6                    & 0\%                   & 0\%                      & 0\%                         & 0\%                          \\ \hline
7                    & 0\%                   & 0\%                      & 0\%                         & 0\%                          \\ \hline
8                    & 0\%                   & 0\%                      & 0\%                         & 0\%                          \\ \hline
9                    & 0\%                   & 0\%                      & 0\%                         & 0\%                          \\ \hline
10                   & 0\%                   & 0\%                      & 0\%                         & 0\%                          \\ \hline
\end{tabular}
\end{table}


\begin{table}[h]
\centering
 \caption{Relative improvement in fault detection effectiveness of the data diversity-based approach compared to random baseline for Convolutional Neural Network}
     \label{tab:RL_method3_random_CNN}
\begin{tabular}{|c|c|c|}
\hline
\textbf{MR Set-Size} & \textbf{Anomaly-based} & \textbf{Clustering-based} \\ \hline
1                    & 2.38\%*                     & 66.66\%*                       \\ \hline
2                    & 1.81\%*                  & 27.27\%*                     \\ \hline
3                    & 1.63\%*                  & 14.75\%*                     \\ \hline
4                    & 1.51\%*                   & 6.06\%*                      \\ \hline
5                    & 1.44\%*                  & 1.44\%*                     \\ \hline
6                    & 1.38\%*                  & -2.77\%*                    \\ \hline
7                    & 1.35\%*                 & 8.18\%*                 \\ \hline
8                    & 1.31\%*                     & 5.26\%*                       \\ \hline
9                    & 1.29\%*                  & 3.89\%*                   \\ \hline
10                   & 1.28\%*                   & 2.56\%*                      \\ \hline
11                   & 0\%                      & 0\%                         \\ \hline
\end{tabular}
\end{table}

\begin{table}[h]
\centering
\caption{Average relative improvement in fault detection effectiveness of data diversity-based approach over statement coverage based approach for IBk}
\label{tab:statementbased_Method3_IBk}
\begin{tabular}{|p{1.2cm}|p{1.2cm}|p{1.2cm}|p{1.2cm}|p{1.2cm}|}
\hline
\textbf{MR Set-Size} & \textbf{Rule-based} & \textbf{Anomaly-based} & \textbf{Clustering-based} & \textbf{Data Distribution} \\ \hline
1                    & 39.02\%*                & 3.20\%*                   & 34.14\%*                         & 14.63*                     \\ \hline
2                    & 36.73\%*                & 26.53\%*                  & 22.44\%*                     & 6.12\%*                     \\ \hline
3                    & 9.37\%*              & 1.56\%*                  & 7.81\%*                    & 10.93\%*                     \\ \hline
4                    & 7.46\%*               & 2.98\%*                  & 5.97\%*                     & 1.49\%*                     \\ \hline
5                    & 8.69\%*                & 7.24\%*                      & 10.14\%*                      & 0\%                      \\ \hline
6                    & 8.33\%*                   & 4.16\%*                      & 6.94\%*                      & 2.77\%*                      \\ \hline
7                    & 9.45\%*                & 2.70\%*                  & 4.05\%*                         & 2.70\%*                      \\ \hline
8                    & 7.89\%*              & 2.63\%*                  & 2.63\%*                     & 2.63\%*                     \\ \hline
9                    & 5.0\%*                   & 1.25\%*                  & -2.50\%*                     & 3.75\%*                      \\ \hline
10                   & 2.43\%*                   & 1.21\%                  & -1.21\%                     & 2.43\%*                          \\ \hline
11                   & 0\%                   & 0\%                      & 0\%                         & 0\%                          \\ \hline
\end{tabular}
\end{table}

\begin{table}[h]
\centering
\caption{Average relative improvement in fault detection effectiveness of data diversity-based approach over branch coverage based approach for IBk}
\label{tab:branchbased_Method3_IBk}
\begin{tabular}{|p{1.2cm}|p{1.2cm}|p{1.2cm}|p{1.2cm}|p{1.2cm}|}
\hline
\textbf{MR Set-Size} & \textbf{Rule-based} & \textbf{Anomaly-based} & \textbf{Clustering-based} & \textbf{Data Distribution} \\ \hline
1                    & 26.66\%*                & 26.66\%*                   & 22.22*                         & 22.22\%*                     \\ \hline
2                    & 36.73\%*                & 26.53\%*                  & 22.44\%*                     & 6.12\%*                     \\ \hline
3                    & 11.11\%*              & 3.17\%*                  & 9.52\%*                    & 9.52\%*                     \\ \hline
4                    & 5.88\%*               & 1.47\%*                  & 4.41\%*                     & 2.49\%*                     \\ \hline
5                    & 5.63\%*                & 4.22*                      & 7.04\%*                      & 2.81\%*                      \\ \hline
6                    & 6.84\%*                   & 2.73\%*                      & 5.47\%*                      & 1.36\%*                      \\ \hline
7                    & 8\%*                & 1.33\%*                  & 2.66\%*                         & 1.33\%                      \\ \hline
8                    & 6.49\%*              & 1.29\%*                  & 1.29\%*                     & 1.29\%                      \\ \hline
9                    & 6.32\%*                   & 2.53\%*                  & -1.26\%                     & 5.06\%*                      \\ \hline
10                   & 2.43\%*                   & 1.21\%*                  & -1.21\%*                     & 2.43\%*                          \\ \hline
11\%                   & 0\%                   & 0\%                      & 0\%                         & 0\%                          \\ \hline
\end{tabular}
\end{table}

\begin{table}[h]
\centering
\caption{Average relative improvement in fault detection effectiveness of data diversity-based approach over statement and branch coverage based approach for Naive Bayes}
\label{tab:statementbased_Method3_NB}
\begin{tabular}{|p{1.3cm}|p{1.3cm}|p{1.3cm}|p{1.3cm}|p{1.3cm}|}
\hline
\textbf{MR Set-Size} & \textbf{Rule-based} & \textbf{Anomaly-based} & \textbf{Clustering-based} & \textbf{Data Distribution} \\ \hline
1                    & 40\%*                  & 42.85\%*                      & 42.85*                         & 42.85\%*                          \\ \hline
2                    & 9.80\%*                   & 9.89\%*                      & 9.89\%*                         & 9.898\%*                          \\ \hline
3                    & 2.04\%*                   & 2.04\%*                      & 2.04\%*                         & 2.04\%*                          \\ \hline
4                    & 1.01\%                   & 1.01\%                      & 1.01\%                         & 1.01\%                          \\ \hline
5                    & 0\%                   & 0\%                      & 0\%                         & 0\%                          \\ \hline
6                    & 0\%                   & 0\%                      & 0\%                         & 0\%                          \\ \hline
7                    & 0\%                   & 0\%                      & 0\%                         & 0\%                          \\ \hline
8                    & 0\%                   & 0\%                      & 0\%                         & 0\%                          \\ \hline
9                    & 0\%                   & 0\%                      & 0\%                         & 0\%                          \\ \hline
10                   & 0\%                   & 0\%                      & 0\%                         & 0\%                          \\ \hline
\end{tabular}
\end{table}

\begin{table}[h]
\centering
\caption{Average relative improvement in fault detection effectiveness of data diversity-based approach over statement and branch coverage based approach for Linear Regression}
\label{tab:stmtbranchbased_Method3_LR}
\begin{tabular}{|>{\centering\arraybackslash}p{1.7cm}|>{\centering\arraybackslash}p{1.7cm}|>{\centering\arraybackslash}p{1.7cm}|>{\centering\arraybackslash}p{1.7cm}|}
\hline
\textbf{MR Set-Size} & \textbf{Anomaly-based} & \textbf{Clustering-based} & \textbf{Data Distribution} \\ \hline
1                    & 33.33\%*                 & 30.43\%*                       & 33.33\%*                      \\ \hline
2                    & 6.71\%*                      & 6.77\%*                      & 8.47\%*                       \\ \hline
3                    & 1.53\%*                  & 1.53\%*                      & 9.23\%*                      \\ \hline
4                    & 1.44\%*                 & 1.44\%*                      & 5.79\%*                      \\ \hline
5                    & 1.38\%*                 & 1.38\%*                         & 13.88\%*                      \\ \hline
6                    & 1.35\%*                 & 1.35\%*                        & 10.81\%*                      \\ \hline
7                    & 0\%                 & 0\%                         & 7.89\%*                          \\ \hline
8                    & 2.59\%*                  & 2.59\%*                         & 6.49\%*                          \\ \hline
9                    & 3.79\%*                      & 5.12\%*                         & 3.79\%*                          \\ \hline
10                   & 2.5\%*                      & 2.5\%*                         & 2.58\%*                          \\ \hline
11                   & 0\%                      & 0\%                         & 0\%                          \\ \hline
\end{tabular}
\end{table}

\begin{table}[h]
\centering
\caption{Average relative improvement in fault detection effectiveness of data diversity-based approach over statement and branch coverage based approach for Convolutional Neural Network}
\label{tab:faultbased_Method3_CNN}

\begin{tabular}{|c|c|c|}
\hline
\textbf{MR Set-Size} & \textbf{Anomaly-based} & \textbf{Clustering-based} \\ \hline
1                    & 86.95\%*                   & 204.38\%*                      \\ \hline
2                    & 1.81\%*                  & 27.27\%*                     \\ \hline
3                    & 0\%                 & 12.90\%*                    \\ \hline
4                    & 1.51\%*                 & 6.06\%*                    \\ \hline
5                    & 0\%                 & 0\%                    \\ \hline
6                    & 1.38\%*                 & -2.77\%                    \\ \hline
7                    & 1.35\%*                 & 8.10\%*                     \\ \hline
8                    & 1.31\%*                 & 5.26\%*                     \\ \hline
9                    & 0\%                 & 2.56\%*                     \\ \hline
10                   & 0\%                      & 1.26\%*                         \\ \hline
11                   & 0\%                      & 0\%                         \\ \hline
\end{tabular}
\end{table}

\begin{table}[h]
\centering
\caption{Average percentage of fault detected for subject programs}
\label{tab:apfd_subjects}

\begin{tabular}{|>{\centering\arraybackslash}p{1cm}|>{\centering\arraybackslash}p{1cm}|>{\centering\arraybackslash}p{1cm}|>{\centering\arraybackslash}p{1cm}|>{\centering\arraybackslash}p{1cm}|>{\centering\arraybackslash}p{1cm}|}
\hline
\textbf{Subject}  & \textbf{Data Distribution} & \textbf{Rule Based} & \textbf{Clustering Based} & \textbf{Anomaly Based} & \textbf{Random Based} \\ \hline
IBk               & 0.85                       & -                   & 0.91                      & 0.72                   & 0.72                  \\ \hline
Linear Regression & 0.89                       & 0.92                & 0.97                      & 0.95                   & 0.78                  \\ \hline
CNN               & -                          & -                   & 0.95                      & 0.95                   & 0.85                  \\ \hline
Naive Bayes       & 0.90                       & 0.93                & 0.95                      & 0.92                   & 0.67                  \\ \hline
\end{tabular}
\end{table}

\subsubsection{Effective number of MRs used for Testing}
In Figure~\ref{Fig:allsubject_methods_killrateMRs}, we show the effective number of MRs using the fault detection thresholds for the subject programs. The pink vertical lines represent effective MR set size when using the rule-based, green vertical line represents data distribution, brown vertical line represents the clustering-based and the orange vertical lines represent the same for the random baseline. The respective thresholds are shown near each vertical line.

 Similarly for CNN, clustering based approach provides a effective MR set size of 2 for the 2.5\% threshold and random approach provides a effective MR set size of 4 for the 2.5\% threshold. In Naive Bayes, anomaly based approach provides a effective MR set size of 1 for the 2.5\% threshold and random approach provides a effective MR set size of 3 for the 2.5\% threshold. In Linear Regression, data diversity based approach provides an effective MR set size of 1 for the 2.5\% threshold and random approach provides a effective MR set size of 7 for the 2.5\% threshold. This practically leads to 85\% reduction in MR set size when compared to random approach. Therefore for all the subject programs, using our proposed methods resulted in reduction in the effective MR set size compared to using a random ordering of MRs. 

\subsubsection{The average time taken to detect a fault } Tables ~\ref{time_method3_random_LR} and~\ref{time_method3_random_IBk} shows the average time taken to detect a fault in Linear Regression and IBk respectively. As shown in these results, using Data Distribution resulted in significant reductions in the average time taken to detect a fault by 26\% when compared to random baseline.  Table~\ref{tab:time_metrics} shows the time taken to generate the metrics for the SUT. We observed that the time taken to generate the metrics for the SUT was less than a minute. Therefore, our proposed approach is time and cost effective when prioritizing MRs.
\subsubsection{Average percentage of fault detected} Table~\ref{tab:apfd_subjects} shows the APFD for the SUT. We can observe that for all our subject programs, our proposed metrics provided higher APFD value between 0.72-0.92 when compared to random based prioritization. As a result, our metrics provides higher rate of fault detection when compared to random approach.

\subsection{RQ2: Comparison of proposed MR Prioritization approach against code coverage-based approach}


In this research question, we evaluate whether the data diversity based approach outperforms code coverage-based approach. In order to answer the research question, we formulate the following hypotheses:

\begin{itemize}
    \item[$H_{2}$:] For a given MR set of size $m$, the fault detection effectiveness of the MR set produced using data diversity based approach is higher than the fault detection effectiveness of the MR set produced by the code coverage-based prioritization.
    \end{itemize}

The null hypothesis $H_{0x}$ for each of the above defined Hypotheses $H_{x}$ is that the MR sets generated by the proposed method performs equal to or worse than the MR sets generated by the code coverage-based prioritization in terms of fault detection effectiveness.

Table~\ref{tab:statementbased_Method3_IBk} and Table~\ref{tab:branchbased_Method3_IBk} shows the relative improvement in fault detection percentage between the MR sets generated by data diversity based approach over statement and branch coverage based approach for IBk. We observed that the rule based, anomaly and clustering based metric provided improvement in fault detection over statement coverage based approach for all the MR set sizes except $m=11$. Therefore we can reject the null hypotheses $H_{02}$. 

Table~\ref{tab:statementbased_Method3_NB} shows the relative improvement in fault detection percentage between the MR sets generated by data diversity based approach and code coverage based approach for Naive Bayes. We observed that all the metric provided improvement in fault detection for MR set size $m=1$ to $m=4$. Therefore we cannot reject the null hypotheses $H_{02}$ in general for Naive Bayes. Table~\ref{tab:stmtbranchbased_Method3_LR} shows the relative improvement in fault detection percentage between the MR sets generated by data diversity based approach over statement and branch coverage based approach for Linear Regression. We observed that anomaly based and data distribution metric provided improvement in fault detection only for all MR set sizes except $m=11$. Therefore we reject the null hypotheses $H_{02}$ in general for Linear Regression. Table~\ref{tab:faultbased_Method3_CNN} shows the relative improvement in fault detection percentage between the MR sets generated by data diversity based approach over statement and branch coverage based approach for CNN. We observed that the clustering-based approach showed improvement in fault detection for all the MR set sizes and anomaly based approach showed improvement in fault detection only for MR set size $m=1$ to $m=5$. Therefore, we reject the null hypothesis $H_{02}$ in general for clustering-based metric.
Overall, our proposed metrics outperforms the statement and branch coverage based approach for the subject programs.

\begin{table}[h]
\centering
\caption{Comparison of average time taken to detect a fault using Data Diversity method and Random baseline for Linear Regression and IBk}
 \label{time_method3_random_LR}
\begin{tabular}{|>{\centering\arraybackslash}p{1cm}|>{\centering\arraybackslash}p{1cm}|>{\centering\arraybackslash}p{1cm}|>{\centering\arraybackslash}p{1cm}|>{\centering\arraybackslash}p{1cm}|>{\centering\arraybackslash}p{1cm}|}
\hline
\textbf{Subject}  & \textbf{Tsp}   & \textbf{Data Distribution} & \textbf{Rule Based} & \textbf{Random Baseline} & \textbf{\% Time Reduction} \\ \hline
IBk               & Test1          & -                          & 8910.80s            & 12562.49s                & 29.06\%                    \\ \hline
Linear Regression & Prediction & 17057.43s                  & -                   & 21609.766s               & 26.06\%                    \\ \hline
\end{tabular}
\end{table}

\begin{table}[h]
\centering
\caption{Comparison of average time taken to detect a fault using Data Diversity method and Random baseline for IBk}
 \label{time_method3_random_IBk}
 \begin{tabular}{|>{\centering\arraybackslash}p{1cm}|>{\centering\arraybackslash}p{1cm}|>{\centering\arraybackslash}p{1cm}|>{\centering\arraybackslash}p{1cm}|>{\centering\arraybackslash}p{1cm}|}
\hline
$\mathbf{{T_{sv}}}$ &\textbf{$\mathbf{{F_v}}$} & \textbf{Rule Based} & \textbf{Random baseline}&\textbf{\% Time reduction} \\ \hline
 Testset1 &    PIT & 8910.80s  & 12562.49s & 29.06\%   \\ \hline
 Testset2 & Major  & 6148.29s & 7508.29s & 22.23\%   \\ \hline
\end{tabular}
\end{table}

\begin{table}[!htbp]
\centering
\caption{Time taken to generate the metrics for SUT}
 \label{tab:time_metrics}
\begin{tabular}{|>{\centering\arraybackslash}p{1.3cm}|>{\centering\arraybackslash}p{1.3cm}|>{\centering\arraybackslash}p{1.3cm}|>{\centering\arraybackslash}p{1.3cm}|>{\centering\arraybackslash}p{1.3cm}|}
\hline
\textbf{SUT}      & \textbf{Rule-based} & \textbf{Anomaly-based} & \textbf{Clustering-based} & \textbf{Data Distribution} \\ \hline
IBk               & 9 sec               & 7 sec                  & 17 sec                     & 6 sec                      \\ \hline
Linear Regression & -               & 4 sec                  & 20 sec                     & 3 sec                      \\ \hline
Naive Bayes       & 8 sec               & 6 sec                  & 18 sec                     & 5 sec                      \\ \hline
CNN               & N/A                 & 200 sec                & 300 sec                   & N/A                        \\ \hline
\end{tabular}
\end{table}

\section{Related Work}
Prior work~\cite{anonymous2021} proposed fault-based and code coverage-based approach to prioritize MRs in the context of regression testing. The fault-based approach applies a greedy algorithm on the faults detected by the MRs on the previous version of the SUT to prioritize MRs for testing the next version of the SUT. Code coverage-based approach uses statements and branches covered by the MRs to prioritize MRs. The proposed approach was applied to three diverse applications and the result indicates that the code coverage based approach did not work effectively for machine learning-based programs. 

Mayer and Guderlei~\cite{mayer2006empirical} conducted an empirical study to evaluate the usefulness of MRs. Based on the experiment result, the authors devise several criteria to asses the quality of the MRs based on the potential usefulness. These criteria were defined based on the different execution paths undertaken by the source and follow-up test cases. While this work provides some directions into selecting useful MRs, they do not directly address MR prioritization. In particular, we focus on developing automated methods for MR prioritization. Hui et al.~\cite{hui2015measurable} provided qualitative guidelines for MR prioritization. The guidelines could be used to select effective MRs and improve the performance of MT. The author also provides three measurable metrics for MR selection: in-degree of MR, algebra complexity of MR, and distance between test inputs of MR. The empirical result shows inconsistency between the correlation value and the metrics, indicating that other factors could influence the effectiveness of MR. Mani et al.~\cite{mani2019coverage} proposed metrics to identify the quality of dataset based on centroid positioning, pairwise boundary conditioning and equivalence partitioning. This work does not provide insight on the diversity in the dataset and does not address MR prioritization. Sun et al.~\cite{SUN2022111091} proposed a technique to generate good source test cases in metamorphic testing based on constraint solvers and techniques of symbolic execution. Also, a prioritization of source test cases was conducted based on path distances among test cases. This work does not focus on prioritization of metamorphic relations. Also, the work does not conduct prioritization of metamorphic relations for machine learning programs.

\section{Threats to Validity}
\label{section:ThreattoValidity}

\textbf{External Validity: }
Our proposed metric is applied to supervised ML classifiers. Metrics such as rule-based is not suitable for unsupervised classifiers and data-distribution metric can be applied only to data set with numerical values. We plan to design more metrics to cover unsupervised classifiers and non-numerical data. In this work, all the experiment subjects are
implementation of machine learning algorithm used in the Weka ML library. Although they are popular algorithms and widely used, our findings may not be generalizable to commercial machine learning projects such as autonomous vehicles. To mitigate this threat, we plan to
explore the effectiveness of our proposed approach on AI systems in the health care domain and Apollo autonomous driving platform projects in the future.

\textbf{Internal Validity: }
We used dataset having an average of 500000 instances and 90 attributes on our subject programs. However, the dataset size could be small for generating the proposed metrics and prioritizing the MRs.  We used less than 100 mutants for validating our prioritized MR ordering for CNN program. However, it is possible that the number of mutants used is low.  

\section{Conclusion}

 Metamorphic relations in metamorphic testing have multiple source and follow-up test cases. As a result, execution and training of machine learning model with large source and follow-up data set can exponentially increase the execution time and cost of metamorphic testing for ML applications. To overcome this problem, we developed four metrics based on diversity between the source and follow-up data set for prioritization of MRs. 

We evaluated our proposed metrics on four open-source machine learning programs. The experimental results show that utilizing MR prioritization on ML applications can increase the fault detection effectiveness of a given MR set up to 33\% for linear regression, 86\% for CNN and up to 40\% for Naive Bayes. In particular, using our proposed MR prioritization approaches reduced the average time taken to detect a fault compared to most common approach of randomly executing the MRs by 30\% for KNN classifier implementation. Further, our results show that data diversity based metrics outperforms code coverage based approach for prioritizing MRs by 40\%. Our proposed metrics also provided higher rate of fault detection when compared to random based prioritization. Our finding also indicates that the number of MRs needed to execute and test ML programs reduced by 66\% which translates to saving cost and time for software testing practitioners.


\clearpage 
\bibliographystyle{IEEEtran}
\bibliography{reference}

\end{document}